Resonance sticking in the scattered disk

Patryk Sofia Lykawka[a,*] and Tadashi Mukai[a]

[a]Kobe University, Graduate School of Science – Department of Earth and Planetary Sciences, 1-1 rokkodai, nada-ku, Kobe, 657-8501, Japan
[*]Corresponding Author e-mail address: patryk@dragon.kobe-u.ac.jp

Pages: 24
Figures: 7
Tables: 1

**Proposed Running head:**
Resonance sticking in the scattered disk

**Editorial correspondence to:**
Patryk Sofia Lykawka
Kobe University
Graduate School of Science
Department of Earth and Planetary Sciences
1-1 rokkodai, nada-ku, Kobe
657-8501
Japan
Phone number: +81 (0)78 803-5740
FAX: +81 (0)78 803-6483
E-mail: patryk@dragon.kobe-u.ac.jp; patryksan@gmail.com

**Abstract**

We investigate the dynamical evolution of trans-Neptunian objects (TNOs) in typical scattered disk orbits (scattered TNOs) by performing simulations using several thousand particles lying initially on Neptune-encountering orbits. We explore the role of resonance sticking in the scattered disk, a phenomenon characterized by multiple temporary resonance captures ('resonances' refers to external mean motion resonances with Neptune, which can be described in the form *r:s*, where the arguments *r* and *s* are integers). First, all scattered TNOs evolve through intermittent temporary resonance capture events and gravitational scattering by Neptune. Each scattered TNO experiences tens to hundreds of resonance captures over a period of 4 Gyr, which represents about 38% of the object's lifetime (mean value). Second, resonance sticking plays an important role at semimajor axes *a* < 250 AU, where the great majority of such captures occurred. It is noteworthy that the stickiest (i.e., dominant) resonances in the scattered disk are located within this distance range and are those possessing the lowest argument *s*. This was evinced by *r*:1, *r*:2 and *r*:3 resonances, which played the greatest role during resonance sticking evolution, often leading to captures in several of their neighboring resonances. Finally, the timescales and likelihood of temporary resonance captures are roughly proportional to resonance strength. The dominance of low *s* resonances is also related to the latter. In sum, resonance sticking has an important impact on the evolution of scattered TNOs, contributing significantly to the longevity of these objects.

# 1. Introduction

Trans-Neptunian objects (TNOs) are the leftovers of a primordial disk of planetesimals in the outer solar system. Thus, they carry precious information about the origin and evolution of the solar system (Luu and Jewitt, 2002; Morbidelli and Brown, 2004). These icy bodies orbit mainly in two reservoirs: the trans-Neptunian belt (or Edgeworth-Kuiper belt), at semimajor axes 30 AU < $a$ < 48 AU, and the scattered disk at typically $a$ > 48 AU[1]. Members of the scattered disk show moderate to large eccentricities $e$, and inclinations, $i$, of up to ~47°. Distinct classes of bodies have also been identified in the scattered disk: scattered, detached and resonant. Scattered TNOs[2] show perihelion distances 30 AU < $q$ < 37-40 AU, whereas detached TNOs have $q$ > 37-40 AU. Detached TNOs never encounter Neptune over 4-5Gyr, thus they appear to be "detached" from the solar system. The observed detached population is more easily identified at $q$ > 40 AU (see Gladman et al., 2002; Lykawka and Mukai, 2007b for details). Moreover, scattered TNOs contribute to the Centaurs, Jupiter-family and Halley-type comets, and the Oort cloud (Duncan and Levison, 1997; Fernandez et al., 2004; Emel'yanenko et al., 2005; Levison et al., 2006). In addition, several of the known TNOs in the scattered disk (~1/3) are also currently situated in resonances with Neptune[3], in the resonances between the 9:4 and the 27:4 (Lykawka and Mukai, 2007a). In a resonant state, a TNO will be protected against encounters with Neptune by the libration mechanism, which tends to maximize the relative distance between the minor body and the giant planet (e.g., Malhotra, 1996).

The origin and orbital evolution of scattered TNOs can be explained by gravitational scattering of planetesimals by Neptune, suggesting that these objects represent ~1% of a yet larger population which existed in the past on Neptune-encountering orbits (e.g., primordial planetesimals at 25-35 AU) (Duncan and Levison, 1997; Morbidelli et al., 2004). Furthermore, Morbidelli (2005) has shown that the observed scattered population cannot originate from the current trans-Neptunian belt, giving strong support for the scenario proposed by Duncan and Levison (1997), as mentioned above. Finally, one should note that this scenario is also valid in various models for the formation of the trans-Neptunian region structure (Morbidelli and Brown, 2004; Hahn and Malhotra, 2005; Chiang et al., 2007; Levison et al., 2007). That is, in these models Neptune migrates outwards to its present orbit during the early solar system, an epoch at which there was still a large remaining population of planetesimals on Neptune-encountering orbits.

Another important (but poorly explored) mechanism dictating the orbital evolution of scattered TNOs is resonance sticking, defined as a single or multiple temporary resonance capture(s) during the object's dynamical lifetime. Jumps between resonances can occur in regions where resonances overlap (Robutel and Laskar, 2001). Resonance sticking was noted by Duncan and Levison (1997) and Gladman et al. (2002) in their simulations of scattered disk objects. Furthermore, resonance sticking could also play a major role in determining the surprisingly high survival rate of scattered TNOs after billions of years (Malyshkin and Tremaine, 1999). In preliminary investigations, Lykawka and Mukai (2004) have shown that resonance sticking is a very common phenomenon among members of the scattered disk. In fact, *all* scattered bodies that evolved over 4 Gyr experienced resonance sticking. Moreover, resonances of the type *r*:1 and *r*:2 played a major role during the evolution (Lykawka and Mukai, 2004; Lykawka and Mukai, 2006; Gallardo, 2006a). As a matter of fact, Gallardo (2006b) has demonstrated that *r*:1 and *r*:2 resonances are the strongest in the scattered disk. Interestingly, these resonances can sometimes promote scattered bodies to the detached population. This particular phenomenon is explained by the Kozai mechanism inside such strong resonances (Gomes et al., 2005; Gallardo, 2006a).

---

[1] The trans-Neptunian belt and the scattered disk have no clear dynamical boundaries. See Morbidelli and Brown (2004), and Lykawka and Mukai (2007b) for details.
[2] Henceforth, 'scattered TNOs' will refer to observed bodies, and 'scattered particles', 'scattered objects' and similar designations will stand for objects from simulations.
[3] For the sake of brevity, we will refer to 'resonance', which we take to mean any external mean motion resonance with Neptune. We will describe these resonances by $r:s$, where the arguments $r$ and $s$ are integers, and define the resonance order as the value given by $r$ minus $s$.

In addition, Fernandez et al. (2004) and Gomes et al. (2005) also present examples of objects that were temporarily captured in distant resonances (> 50 AU). In particular, the reported objects spent quite long time intervals locked in strong resonances (e.g., *r*:1 resonances). On the other hand, we stress that the resonance sticking phenomenon represents the influence of resonances on the dynamical evolution of scattered TNOs over the age of the solar system. That is, in principle a scattered body can experience capture events in *any* resonance at a given time, presumably totaling several distinct resonances during its lifetime. Therefore, the role of resonance sticking in the scattered disk *per se* has not been explored in previous works (except Lykawka and Mukai, 2006). Indeed, the examples discussed in those publications present an extremely small fraction of all resonance capture events that each TNO can experience in the scattered disk.

Here, we explore the origin and dynamical evolution of bodies in the scattered disk by means of numerical simulation, aiming to further explore the resonance sticking phenomenon over 4 Gyr. We provide extensive information on the captured resonant objects from the scattered disk, such as the resonances occupied, timescales, resonance strengths, among others. Furthermore, the sample used in our simulations is much larger than that in previous published studies of long-term dynamical evolution of scattered bodies. Consequently, we obtained a larger surviving population at the end of the simulations (i.e., better statistics).

## 2. Numerical methods

We numerically evolved a population of 22380 particles in orbits that were initially Neptune-encountering, with initial 30 AU < *a* < 50 AU, 25 AU < *q* < 35 AU and *i* up to 20° (uniform distributions). These initial conditions were chosen to focus on the origin and further evolution of scattered TNOs after the end of planet migration (e.g., Morbidelli and Brown, 2004), in line with models for the origin of this population (see Section 1). We followed the system until 4 Gyr using the hybrid symplectic integrator EVORB, with a time step of 6 months (e.g., Brunini and Melita, 2002; Fernandez et al., 2002). Orbital elements refer to the heliocentric frame. The giant planets were fully considered as massive perturbers in a self-consistent way, and the minor bodies were treated as massless particles suffering perturbations from just the massive bodies. Collisions and close encounters with the planets and the Sun were completely taken into account. Particles that collided with a massive body were removed from the integration. The output data was recorded every 2000 yr.

In an *r:s* resonance, the principal resonant angle, $\phi = r\lambda - s\lambda_N - (r-s)\cdot\varpi$, and its amplitude, $A_\phi$, are useful quantities in the study of resonant motion, where *r* and *s* are integers, $\lambda$ and $\lambda_N$ are the mean longitudes of the body and Neptune, and $\varpi$ is the longitude of perihelion of the body. Smaller values of $A_\phi$ indicate larger relative distances for Neptune-minor body encounters (e.g., Murray and Dermott, 1999). We employed our RESTICK code for the identification of resonance capture events and to calculate $A_\phi$ for each resonance detection (Lykawka and Mukai, 2007a). Basically, RESTICK reads the entire output data for an object and searches for resonances in semimajor axes using 125 data point windows – 250 kyr in our simulation. The code is able to identify symmetric and asymmetric behaviors[4]. After visual inspection of a large random sample (tens of objects), we estimate that RESTICK provided reliable resonance identifications for resonant capture durations, $t_{res}$, longer than 275 kyr and $A_\phi$ < 150-170° (< 60° in case of asymmetric behavior) at > 99% confidence level. Notably, false resonance identifications were << 1% for slightly longer values of $t_{res}$ (> 300-400 kyr). Therefore, we were able to follow the entire evolutionary orbital history of each scattered object using the output data from EVORB (scattering by Neptune) and the resonance capture identifications from RESTICK (resonance sticking)[5].

## 3. Results and discussion

Of the 22380 particles, 255 were able to remain in the scattered disk ($a$ < 1000 AU) for the entire 4 Gyr of the simulation, representing about 1% of the initial sample. Final orbital distributions were confined to $i$ < 42° and $q$ < 43 AU (except one object with $q$ ~ 61 AU. See Section 3.1). These results are in agreement with those presented by Duncan and Levison (1997) and Morbidelli et al. (2004). The depletion of scattered particles with time also agrees with a non-random walk approximation, following $\sim t^{-1.5}$ during the last 1.5 Gyr (Malyshkin and Tremaine, 1999). Since our results are compatible with scattered TNOs, we believe the behavior of these bodies can be well described by this work. Henceforth, we will focus our analysis on the evolution of these 255 objects using their orbital elements averaged over the last 100 Myr of simulation.

### 3.1 Resonance sticking: general behavior and individual cases

Firstly, all particles experienced multiple resonance trapping events during their lifetimes, being typically captured tens or hundreds of times in various resonances across the scattered disk. In general, scattered particles were captured on the average in 88 distinct resonances. During non-resonant evolution, the particles suffered encounters with Neptune. Thus, although the dynamical evolution of a scattered object was chaotic in nature, in general it was composed of intermittent phases of gravitational scattering and temporary resonant capture. To better understand this behavior, we present, in Fig. 1, the orbital evolution of some individual examples (Particles 1 to 6, from top to bottom). From Fig. 1, one can visualize scattering and resonance sticking by random walks and periods spent at nearly constant $a$, respectively (right panels). Both phenomena are also visible as widely spaced and highly concentrated dots in element space ($a$-$e$) (left panels).

---
[4] In *r*:1 resonances (e.g., 2:1, 3:1, etc.), both symmetric and asymmetric resonant configurations are possible. See Gallardo (2006b) for details.
[5] For the output data step used in this work, the RESTICK code can identify capture events in any resonance with order ≤ 200 beyond 30 AU over the entire 4 Gyr. There is no limitation in the maximum distance for resonance identification.

In Fig. 1, we indicate several of the resonances where the 6 particles experienced the longest temporary captures. It can be seen that these resonances played a major role in the orbital evolution of the object during resonance sticking. Notice that each object experienced captures in many other resonances. However, we do not show them for clarity of the figure. For example, we detected hundreds of temporary captures in 85 distinct resonances for Particle 1 over the 4 Gyr of the simulation[6] (Fig. 1a). This particle started near the 2:1 resonance, becoming initially trapped there for ~160 Myr. After being scattered, the object spent more than 1 Gyr interacting with, and being captured in, the 17:7, 22:9 and other neighboring resonances. After a short stay in the 16:7 resonance (~35 Myr), the particle was scattered and quickly captured over several tens of Myr in each of the 17:6, 28:9, 19:6, 10:3, 17:5 and other resonances around the same region. At ~3.8 Gyr the particle started to suffer strong encounters with Neptune. Even though several short resonance captures were detected, Particle 1 suffered scattering during the last ~3 Myr, ending with $a$ = 101.3 AU, $q$ = 32.4 AU and $i$ = 11.5°. Overall, Particle 1 spent about 30% of its lifetime trapped in resonances. In case of Particle 2, one can see a remarkable example of the strength of a single resonance. After short resonant captures between 60–90 AU (Fig. 1b, left panel), the object spent ~3.5 Gyr under the control of the 6:1 resonance, remaining locked there until the end of the simulation, at 4 Gyr, with a final $q$ = 61.2 AU. Such a large perihelion was due to the Kozai mechanism inside the 6:1 resonance, with the object becoming temporarily detached from the solar system.

Other outcomes are possible. Particle 3 is a good example of the case where resonance sticking can occur at quite large semimajor axes (Fig. 1c). The evolution of this object was strongly influenced by $r$:1 and $r$:2 resonances, in particular the 23:2, 14:1, 29:2 and 21:1 resonances, with capture timescales around ~100 Myr each. Particle 3 remained locked in the 29:2 resonance for the last 20 Myr of the simulation. The strength of $r$:1 and $r$:2 resonances was also evident in the case of Particle 4, where the 4:1, 9:2, 13:2, 9:1 and 19:2 resonances played a major role (Fig. 1d). In particular, the stay in the latter resonance was ~600 Myr long, a period during which large $q$ and $i$ excursions can be clearly seen. Furthermore, this object also followed temporarily a Centaur-type orbit (e.g., Horner et al., 2003), suffering encounters with Neptune and Uranus at the beginning of the integration. This suggests that some Centaurs could find a way to become scattered TNOs with the help of resonances. Indeed, this idea is strengthened by the behavior of Particle 5, which started in a typical Centaur-type orbit (initial $q$ = 25.3 AU), but quickly evolved to the scattered disk acquiring $q > 30$ AU (Fig. 1e). After short captures in the 13:5, 8:3 and other resonances, the object's evolution was controlled by a group of resonances near the 13:1 resonance, whose combined action stuck the object around $a = 160$-$165$ AU for almost 1.5 Gyr. Finally, the object was temporarily captured in the 17:2 (~150 Myr) and 8:1 (~350 Myr) resonances.

---

[6] Individually, the number of detected distinct resonances over 4 Gyr varied between 4 and 199 among the 255 particles.

Our last example is Particle 6, shown in Fig. 1f. The orbital history of this object not only reinforces the importance of $r$:1 and $r$:2 resonances, but also demonstrates the possibility of two adjacent $r$:1 resonances acting to completely determine the evolution of an object through the secondary action of combined $r$:2, $r$:3, ... resonances, within the location of consecutive $r$:1 resonances. For example, at around 2.5 Gyr the 12:1 and 13:1 resonances dictated the evolution: temporary captures occurred at both resonances and also at several intermediate resonances (37:3, 25:2, 38:3, etc). Note that 25:2 is the lowest order resonance in between 12:1 and 13:1, using the Farey sequence (i.e., (12+13):(1+1) = 25:2. See Hardy and Wright, 1988). Conversely, between 12:1 and 25:2, we find the 37:3 resonance, and so on. In fact, we detected captures in tens of such combined resonances between 12:1 and 13:1, all of them described by the Farey sequence. Nevertheless, we plotted only the resonances with capture timescales of tens to 100-200 Myr in Fig. 1f. Finally, a very similar behavior also occurred for captures around 200 AU, with the action of the 16:1 and 17:1 resonances.

The examples shown in Fig. 1 demonstrate some of the innumerable evolutionary paths of scattered TNOs (see Fernandez et al., 2004; Gomes et al., 2005; Lykawka and Mukai, 2006 for more cases). Although not exhaustive, from these examples and the remaining 249 particles we noticed some typical behaviors: i) resonance sticking involving several neighboring resonances is quite common (around the same region); ii) $r$:1 and $r$:2 resonances play a major role in the resonant evolution, confirming our early investigations (Lykawka and Mukai, 2004; Lykawka and Mukai, 2006), and the expectations of Gallardo (2006a). In addition, $r$:3 resonances were also relatively important during the evolution (Fig. 2); iii) resonance sticking is unimportant beyond about $a = 250$ AU; iv) resonance trapping evolution is associated with larger $q$. That is, resonant objects tended to exhibit larger perihelia when compared to those not in resonances.

3.2 Global analysis of the resonance sticking phenomenon

Based on the RESTICK data compiled from the 255 particles, we identified all detectable resonance captures and obtained their durations over the 4 Gyr available for each particle. In this sample, we found temporary captures in a total of more than 600 distinct resonances, but we limit the discussion below to the 464 resonances that had cumulative total residence timescales greater than ~0.5 Myr.

To start with, we determined the ratio of total time spent locked in resonances to the total dynamical lifetime (4 Gyr), $f_{res}$, for each object. Among the particles, this resulted in fractions varying from a few percent to more than 90% (Fig. 3). Taking into account the entire sample, we found that scattered particles spent on average ~38% of their lifetimes trapped in resonances. This is consistent with the fraction of bodies locked in resonances after 4 Gyr (83 out of 255 objects, or about 33% – Table 1). These fractions should be considered lower limits, however, because RESTICK was unable to identify resonance captures with $t_{res}$ < 250 kyr. In addition, depending on the resonance, the code also failed to identify resonances with $A_\phi$ > 150-170°. We also found a tendency of increasing $f_{res}$ with larger perihelia, finding values of approximately 34, 37 and 55% for $q$ < 35 AU, 35 AU < $q$ < 40 AU and $q$ > 40 AU, respectively. No other apparent correlations of $f_{res}$ and/or $A_\phi$ with orbital elements were found.

In general, resonance sticking occurred mostly at $a$ < 250 AU, corresponding to 99% of the time of all resonance captures in the sample (Fig. 4). This arises for two reasons. Firstly, the probability of capture in resonances at $a$ < 250 AU is higher, since scattered bodies normally need to transverse this inner region before reaching greater distances. Second, the resonances beyond 250 AU are much weaker than those in the inner region (see below and fig. 7 of Gallardo, 2006b).

Furthermore, we determined the importance of each of the 464 resonances during resonance sticking, which we define as 'resonance stickiness'. We calculated the ratio of cumulative total residence time in each resonance to the cumulative total time in all resonances. The number of different particles that were trapped in each resonance was accounted for in this calculation. We set relative resonance stickiness equal to 1 for the 6:1 resonance, which yielded the largest value. We determined the relative resonance stickiness of detected resonances at $a \geq 47.8$ AU (2:1 resonance and beyond) as a function of resonance order ($r - s$) and argument $s$ (Fig. 5). In Fig. 5, the curves in the top panel are plotted for orders up to 25, while those in the bottom panel represent resonances with argument $s$ varying from 1 to 12 ($r$:1, $r$:2, …, $r$:12). Figure 5 shows that resonances of lower order (and lower $s$) have larger resonance stickiness, so scattered TNOs should preferentially be captured and stay longer in these resonances. This is in agreement with the idea that such resonances are dynamically active and possess non-negligible influence in element space ($a$-$e$) (i.e., large resonance widths) (Morbidelli et al., 1995; Malhotra, 1996; Robutel and Laskar, 2001). In any case, we can also conclude from the bottom panel of Fig. 5 that resonances with lower $s$ dominate the resonant evolution in the scattered disk. It can also be seen that resonance stickiness drops rapidly with distance from the Sun, so that only resonances with $s$ = 1, 2 or 3 are relevant beyond 200 AU. The shallower slopes of the $r$:1 and $r$:2 resonances at such large distances are the result of resonant interactions, characterized by irregular short periods of slow circulation and libration of the resonant angle.

To further understand the physical meaning of resonance stickiness and its correlations with distance, resonance order and the argument $s$, we calculated the strength of resonances across the trans-Neptunian region by solving the strength function $SR(a,e,i,\omega) = <R> - R_{min}$, where $\omega$ is the argument of perihelion, $<R>$ is the averaged value of the resonant disturbing function $R$, and $R_{min}$ is the minimum value of $R$ (see Gallardo, 2006b for details). Objects trapped in resonances with high $SR$ are expected to suffer stronger dynamical effects during their evolution. The averaged expansion of $R$ follow $\sim e_N^{j_3} e^{j_4} S_N^{j_5} S^{j_6} [\alpha f_d(\alpha) + f_i(\alpha)] \cos\phi$, where $S = \sin(i/2)$, $\alpha = a_N / a$, and the $N$ subscript refers to Neptune. The expressions $f_d(\alpha)$ and $f_i(\alpha)$ are functions of Laplace coefficients (e.g., Murray and Dermott, 1999; Gallardo, 2006a). In our calculation, we used orbital elements typical of observed scattered TNOs: $q = 33$ AU, $i = 17°$, and $a =$ the nominal semimajor axis of the resonance ($r$:$s$) of interest, $a_{res}$, given by $30.11 \cdot (r/s)^{2/3}$. We set $\omega = 60°$ for all cases.

The strengths of the 464 resonances, calculated in this way, are shown according to resonance order ($r - s$) and argument $s$ in Fig. 6. Note the dependence of the resonant disturbing function on the term $e^{j_4}$, which can be approximately described as $e^{r-s}$ for low inclinations (Gallardo, 2006a). Indeed, resonance order appears in the equations of resonant motion, so the general trend of the stronger resonances having lower orders is not surprising. However, resonance strength is very dependent on its location ($a_{res}$). Indeed, sometimes resonances of high order can be considered equally strong, or even stronger, than other resonances of low order. This occurs because to conserve $q = 33$ AU, the eccentricities in farther resonances become higher, thus implying in stronger resonances following the term $e^{r-s}$. In addition, considering only the strongest resonances of a given order, notice that they become significantly less effective at distances beyond ~150-200 AU (Fig. 6, top panel), which agrees with the fact that resonance captures in our simulation occurred at smaller semimajor axes (Fig. 4). In the bottom panel of Fig. 6, one can easily compare the strength of any particular resonance with the other resonances nearby, for a given distance from the Sun. Another merit is to see the evident drop of strength for $r$:1, $r$:2, ... resonances as a function of their distance from the Sun. Finally, the importance of resonances is limited to a critical semimajor axis, which is smaller for higher values of the argument $s$. Worth noting, extremely high order resonances are found near these critical semimajor axes (recall that $r - s$ increases monotonically with larger $a_{res}$ for a fixed argument $s$), therefore leading to vanishingly small dynamical effects after integrating the equations of motion. For instance, the role of $r$:12 resonances ($s = 12$; Lightest gray curve in Fig. 6, bottom panel) should be negligible at $a > 120$ AU, a region where these resonances reach order $\geq 85$.

We found a very small relative contribution of resonances beyond 250 AU ($r$:1 and $r$:2 type only), corresponding to ~1% of the time of all resonance captures in the simulation (see Fig. 5). In general, we also found $t_{res} < 1$ Myr at those very distant resonances. In addition, Fig. 4 and 5 suggest that the region at ~300-350 AU delimit the outer boundary for resonance sticking in the outer solar system. This agrees with that expected from theory. That is, despite high eccentricities, resonance orders are quite high ($> 30-70$) and the expression ($\alpha \cdot f_d(\alpha) + f_i(\alpha)$) yields vanishingly small contribution (it decreases with larger semimajor axes).

The resemblance between the curves of relative resonance stickiness (Fig. 5) and of resonance strength (Fig. 6) is evident. In this manner, resonance stickiness is related to resonance strength, and vice-versa. In other words, longer temporary captures and higher capture probabilities during resonance sticking are associated with resonance strength. This is in agreement with the idea that scattered bodies should preferentially be captured in stronger resonances, because the latter possess wider resonance widths and penetrate to lower eccentricities. Besides, longer resonant captures are probably connected to stronger dynamical effects experienced by bodies trapped in those resonances (e.g., the resonant motion is more difficult to be disrupted by scattering by Neptune) and to the slow diffusion character at larger perihelia (Murray and Dermott, 1999; Robutel and Laskar, 2001; Gallardo, 2006a; Gallardo, 2006b). In conclusion, we have possible answers to several features discussed previously: i) the ubiquitous contribution of resonances at $a < 250$ AU during resonance sticking reflects the presence of sufficiently strong resonances; ii) the stickier and stronger resonances in the scattered disk have lower $s$. This would naturally explain the dominance of $r$:1, $r$:2 and $r$:3 resonances (Fig. 2; Fig. 5, bottom panel), and also multiple captures in combined resonances following the Farey sequence, since the latter provides resonances with increasing $s$ which lie between two $r$:1 resonances (Fig. 1); iii) the increase of $q$ is associated with longer resonance residence time (i.e., higher resonance stickiness); iv) considering our final $q$-distribution of scattered bodies, and recalling that longer captures are linked to resonance strength, we can say that the $q$-lifting mechanism for the temporary promotion of scattered to detached objects should be important only for $r$:1, $r$:2 and $r$:3 resonances located at about $a < 250$, 200 and 170 AU, respectively. This was observed in our simulations (Fig. 1b,d,e and Fig. 7) and in previous studies (Fernandez et al., 2004; Gomes et al., 2005; Gallardo, 2006a).

## 4. Summary

We performed numerical integrations of an ensemble of particles in Neptune-encountering orbits, following the system until 4 Gyr to better understand the dynamical evolution and the role of resonance sticking in the scattered disk. In summary, other than for a few 4 Gyr-resonant and detached TNOs (Lykawka and Mukai, 2007b and references therein), it seems that all members of the scattered disk have been experiencing resonance sticking with dynamical evolutions in agreement with those described in the present work (see also Lykawka and Mukai, 2006).

The main results of this paper are summarized below:

- The evolution of scattered TNOs is described by multiple temporary resonance locking (resonance sticking) and continuous scattering by Neptune. In general, resonance captures occur tens to hundreds of times, with a total mean timescale representing about 38% of the object's lifetime;

- Resonance sticking occurs for all scattered TNOs and is relevant mostly at $a < 250$ AU. This region contains all resonances with sufficiently high stickiness/strength;

- Resonances (described as $r$:$s$) with the smallest $s$ are the stickiest/strongest in the scattered disk. Therefore, these resonances dominate the region, having longer resonant timescales and higher capture probabilities. In particular, $r$:1, $r$:2 and $r$:3 resonances play the greatest role;

- The temporary promotion of scattered bodies to the detached population ($q$-lifting mechanism) is likely to occur in $r$:1, $r$:2 and $r$:3 resonances located at about $a < 250$, 200 and 170 AU, respectively;

- Timescales and likelihood of temporary resonance captures are roughly proportional to resonance strength, as evinced by the link between resonance stickiness and the latter.

**Acknowledgments**

We would like to thank T. Gallardo and an anonymous referee for useful comments/suggestions that greatly improved this paper. We also thank J. Horner for reading carefully the original manuscript. This study was supported by "The 21st Century COE Program of the Origin and Evolution of Planetary Systems" of the Ministry of Education, Culture, Sports, Science and Technology, Japan.

**Table 1**

**Objects locked in resonances with Neptune in the scattered disk after 4 Gyr**[*]

| $a_{res}$ (AU) [a] | Resonance | $A_\phi$ (°) [b] | Timescale (Myr) [c] |
|---|---|---|---|
| 43.7 | 7:4 | 64±5 | 3000 |
| 46.5 | 23:12 | 80±10 | <10 |
| 49.6 | 19:9 | 34±20 | <1 |
| 51.7 | 9:4 | 44±5 | 3000 |
| 52.0 | 25:11 | 145±20 | <1 |
| 53.0 | 7:3 | 120±5 | 2000 |
| 53.0 | 7:3 | 149±5 | 10 |
| 54.0 | 12:5 | 130±5 | 2000 |
| 55.5 | 5:2 | 157±5 | 2000 |
| 55.5 | 5:2 | 139±5 | 3000 |
| 57.3 | 21:8 | 84±5 | 400 |
| 57.9 | 8:3 | 154±5 | 2000 |
| 57.9 | 8:3 | 107±5 | 2000 |
| 57.9 | 8:3 | 147±5 | 300 |
| 58.6 | 19:7 | 88±5 | 200 |
| 58.8 | 30:11 | 156±20 | <1 |
| 59.1 | 11:4 | 127±5 | 100 |
| 59.8 | 14:5 | 88±5 | 40 |
| 60.6 | 20:7 | 88±5 | 100 |
| 60.6 | 20:7 | 150±20 | 1000 |
| 61.6 | 38:13 | 62±10 | <2 |
| 62.6 | 3:1 | 161±5 | 4000 |
| 62.6 | 3:1 | 161±5 | 3000 |
| 65.4 | 16:5 | 96±5 | 100 |
| 65.4 | 16:5 | 108±5 | 1 |
| 66.1 | 13:4 | 145±5 | 700 |
| 66.1 | 13:4 | 119±5 | 100 |
| 66.1 | 13:4 | 134±5 | 2000 |
| 67.2 | 10:3 | 70±5 | 3000 |
| 67.2 | 10:3 | 118±5 | 1000 |
| 69.4 | 7:2 | 160±5 | 600 |
| 69.4 | 7:2 | 98±5 | 4000 |
| 70.7 | 18:5 | 125±5 | 10 |
| 71.6 | 11:3 | 148±5 | 100 |
| 73.3 | 19:5 | 78±5 | 10 |
| 73.3 | 19:5 | 114±5 | 70 |
| 74.1 | 27:7 | 88±5 | 50 |
| 75.9 | 4:1 | 145±5 | 600 |
| 75.9 | 4:1 | 159±5 | 3000 |
| 75.9 | 4:1 (a90) | 53±10 | 400 |
| 79.0 | 17:4 | 88±5 | 500 |
| 80.0 | 13:3 | 124±5 | 100 |
| 82.9 | 32:7 | 122±5 | 100 |
| 84.1 | 14:3 | 120±5 | 10 |
| 84.1 | 14:3 | 155±20 | 5 |
| 88.0 | 5:1 | 155±5 | 4000 |
| 91.0 | 21:4 | 115±5 | <2 |
| 91.9 | 16:3 | 84±5 | 700 |
| 91.9 | 16:3 | 75±5 | 1 |
| 93.8 | 11:2 | 138±5 | 1000 |
| 95.7 | 17:3 | 102±5 | <2 |
| 98.2 | 53:9 | 92±10 | <1 |
| 99.4 | 6:1 | 135±5 | 3000 |
| 99.4 | 6:1 | 167±5 | 300 |
| 99.4 | 6:1 | 152±20 | 600 |
| 99.4 | 6:1 (a270) | 53±10 | 4000 |
| 103.1 | 19:3 | 107±5 | 800 |
| 110.2 | 7:1 | 164±10 | 1 |
| 110.2 | 7:1 | 168±10 | <1 |
| 110.2 | 7:1 | 166±5 | 2000 |
| 110.2 | 7:1 | 153±5 | 500 |
| 110.2 | 7:1 | 153±5 | 200 |
| 117.6 | 54:7 | 109±5 | 10 |
| 119.3 | 71:9 | 117±20 | 20 |
| 122.9 | 33:4 | 97±5 | 10 |
| 127.0 | 26:3 | 129±5 | 3 |
| 127.0 | 26:3 | 140±5 | 2 |
| 130.3 | 9:1 | 160±5 | 70 |
| 130.3 | 9:1 (a90) | 45±10 | 30 |
| 136.6 | 29:3 | 85±5 | 10 |
| 139.8 | 10:1 | 161±5 | 1 |
| 144.4 | 21:2 | 149±5 | 2000 |
| 145.3 | 53:5 | 123±5 | 3 |
| 145.9 | 32:3 | 140±5 | 40 |
| 148.9 | 11:1 | 165±5 | 1000 |
| 148.9 | 11:1 | 150±5 | 2000 |
| 151.9 | 34:3 | 87±5 | 8 |
| 153.4 | 23:2 | 101±5 | 2000 |
| 153.4 | 23:2 | 135±5 | 3 |
| 174.9 | 14:1 (a90) | 40±10 | 200 |
| 174.9 | 14:1 (a270) | 40±10 | 200 |
| 179.0 | 29:2 | 125±5 | 20 |
| 229.2 | 21:1 | 121±20 | 4 |

[a] Nominal location of a resonance in semimajor axis.

[b] Amplitude of the resonant angle (see text), calculated during a timespan never greater than the last 200 Myr. Errors come from uncertainties in RESTICK calculations. In case of asymmetric behavior ($r$:1 resonances, where $r$ is an integer), the asymmetric center is given in the 'Resonance' column, where '(a90)' and '(a270)' refer to 90 and 270° centers, respectively.

[c] The timescale tells the approximate total time in which the object is inside the resonance (resonance capture duration + interactions).

* Due to the temporary and chaotic character of resonance captures, this list represents just one of uncountable evolutionary outcomes at a given instantaneous time (4 Gyr here). Moreover, a larger initial sample would also yield a larger number of resonant particles, increasing the number of distinct occupied resonances as well.

**Figure captions**

Figure 1. Resonance sticking in the scattered disk: some individual examples (Particles 1-6: panels a-f). We used osculating orbital elements in the heliocentric frame. Left: small dots represent the orbital evolution plotted every 0.5-1 Myr for a period of 4 Gyr. Dashed vertical lines refer to the location of relevant resonances. Perihelia of 30 and 35 AU are shown by two curves (upper and lower, respectively). All particles started on Neptune-encountering orbits, and with $a < 50$ AU. The final orbital elements of the particle are indicated by gray circles (out of range in panel a). Right: evolution of semimajor axis $a$, perihelion distance $q$ and inclination $i$ with time.

Figure 2. Relative importance of resonances (described in the form $r:s$) during resonance sticking for 255 particles that remained in the scattered disk after 4 Gyr. First, we determined the ratio of the total time spent in a particular type of resonance ($s = 1, 2, ..., 12$) to the total time in all resonances. This quantity was later weighted by the number of individual resonances for a given $s$, and normalized to the value found for $r:1$ resonances.

Figure 3. Distribution of the total time spent in resonances to the total dynamical lifetime of 255 particles that remained in the scattered disk after 4 Gyr. See text for discussions.

Figure 4. Cumulative time spent in resonances during resonance sticking and semimajor axes. The time spent in resonances within 150 AU, 200 AU, and 250 AU represents approximately 90%, 97%, and 99% of that computed for all resonance captures experienced by the 255 particles over 4 Gyr, respectively.

Figure 5. Relative resonance stickiness of relevant resonances (described as $r:s$) detected in the evolution of 255 particles that remained in the scattered disk after 4 Gyr. Resonance stickiness illustrates the likelihood of capture into a resonance, and the ability of that resonance to retain a captured object (i.e., timescale) (see text for details). Resonance stickiness was normalized to the largest value, found at the 6:1 resonance ($a = 99.4$ AU). Resonances at $a < 47.8$ AU were omitted due to their overlap with the initial conditions, not allowing a proper calculation of relative resonance stickiness for these resonances. Relative resonance stickiness is given as a function of resonance order (top) and argument $s$ (bottom). We show resonances up to 25th order and with argument $s$ ranging from 1 to 12, with values increasing from black to light gray.

Figure 6. Resonance strength of relevant resonances (described as $r:s$) detected in the evolution of 255 particles that remained in the scattered disk after 4 Gyr. Resonance strength was calculated by solving the *SR* function (see the main text, and Gallardo, 2006b) and it is given as a function of resonance order (top) and argument $s$ (bottom). We show resonances up to 25th order and with argument $s$ ranging from 1 to 12, with values increasing from black to light gray. Compare with Fig. 5.

Figure 7. Orbital distribution of 255 particles that remained in the scattered disk after 4 Gyr (black circles). The orbital elements (heliocentric frame) were averaged over the last 100 Myr. From the left, dashed vertical lines refer to the location of the 10:3, 6:1, 13:2, 7:1, 8:1, 21:2, 11:1, 23:2, 27:2, and 14:1 resonances. These resonances played an important role in lifting the perihelion of some objects.

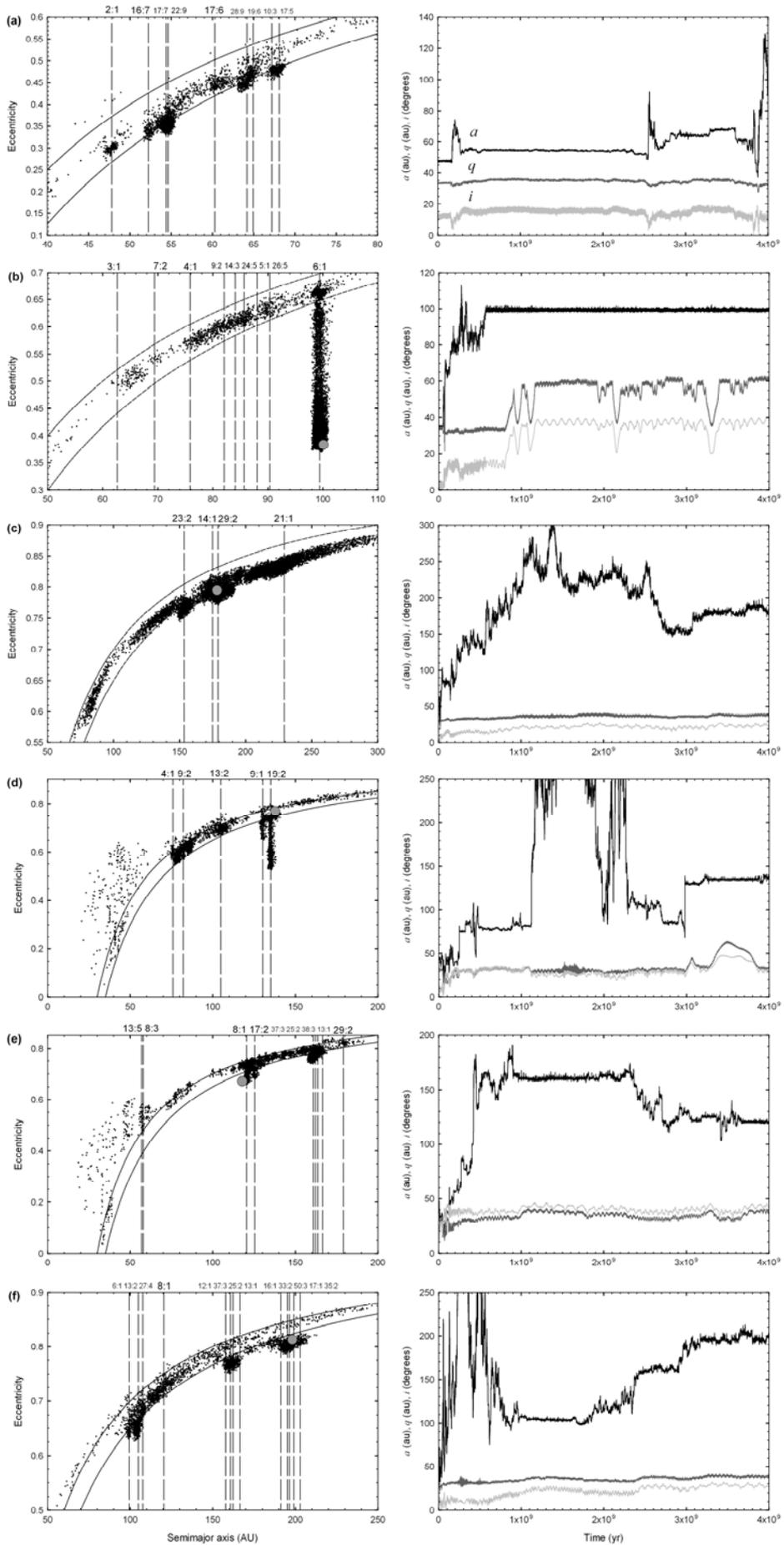

**Figure 1 – Resonance sticking in the scattered disk..., Lykawka, P. S.**

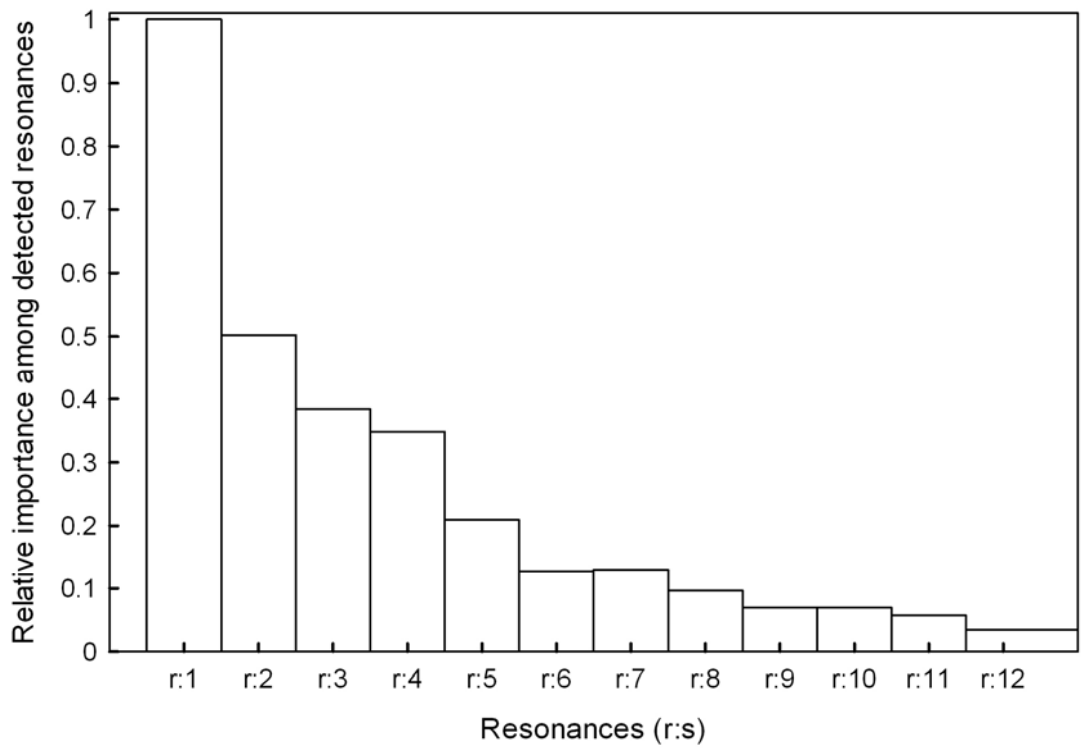

**Figure 2 – Relative importance of r:s resonances as a group..., Lykawka, P. S.**

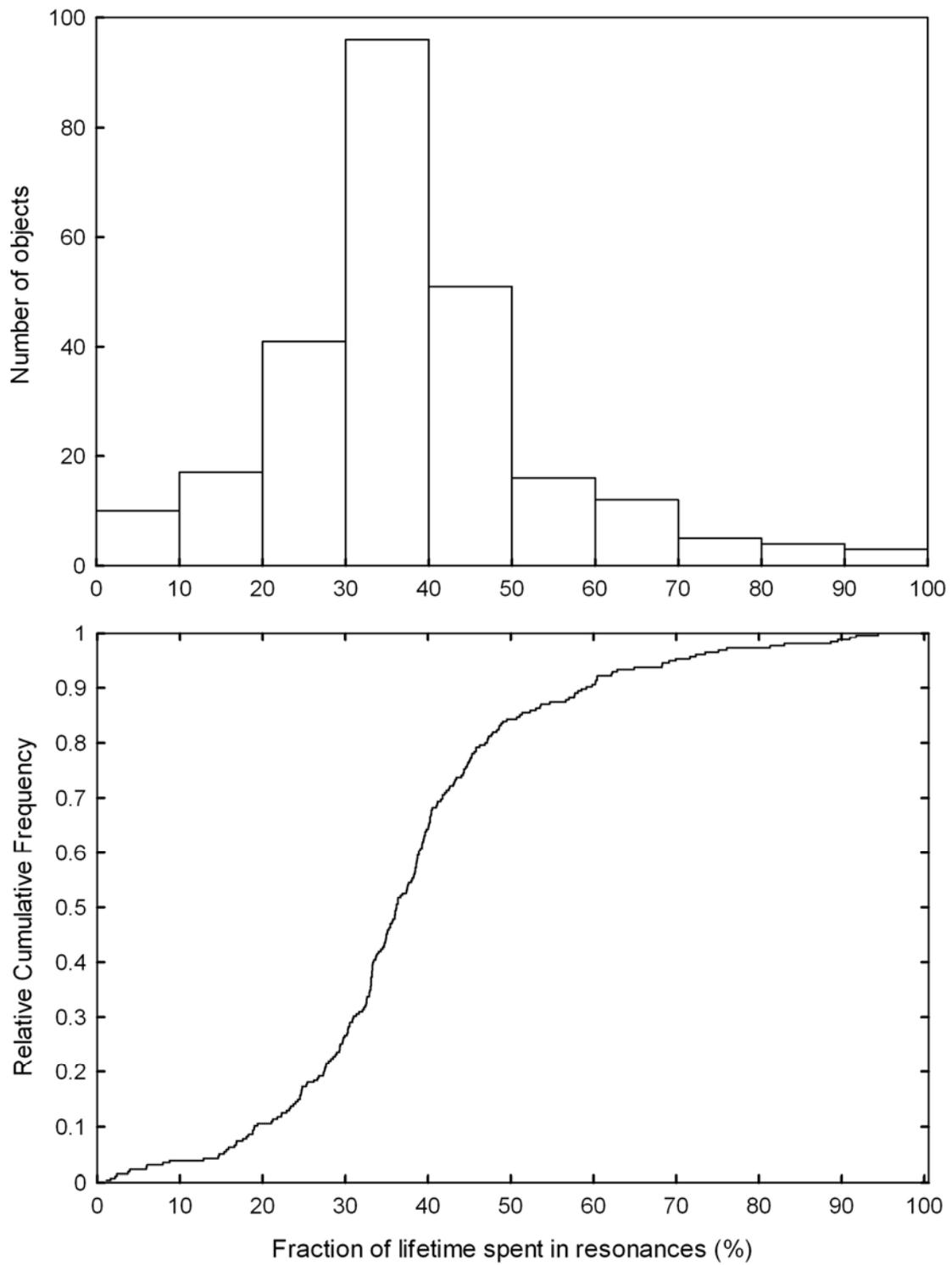

**Figure 3 – Distribution of residence time in resonances..., Lykawka, P. S.**

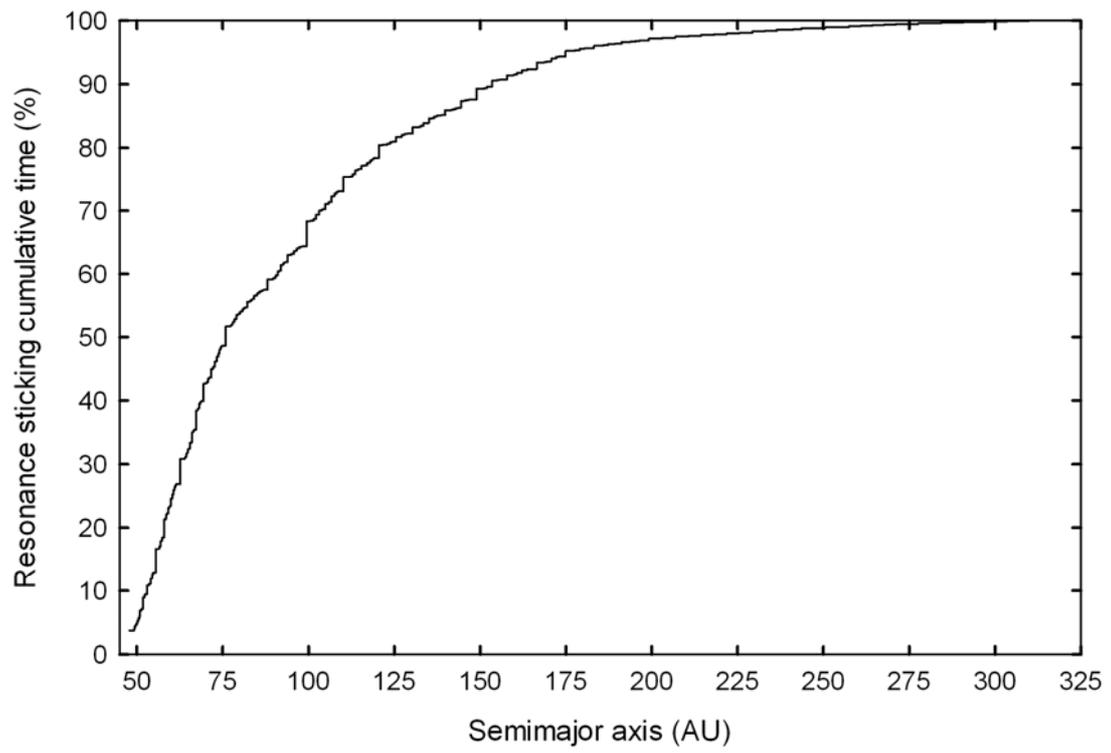

**Figure 4 – Resonance sticking and dependence on semimajor axis..., Lykawka, P. S.**

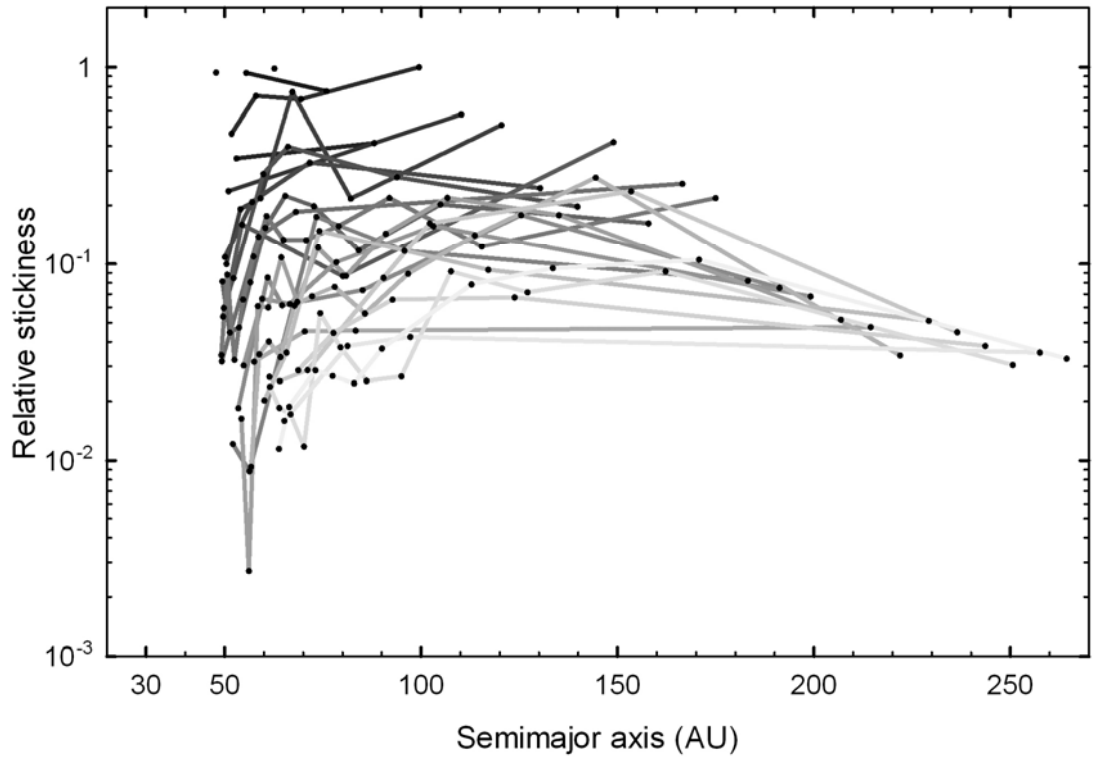

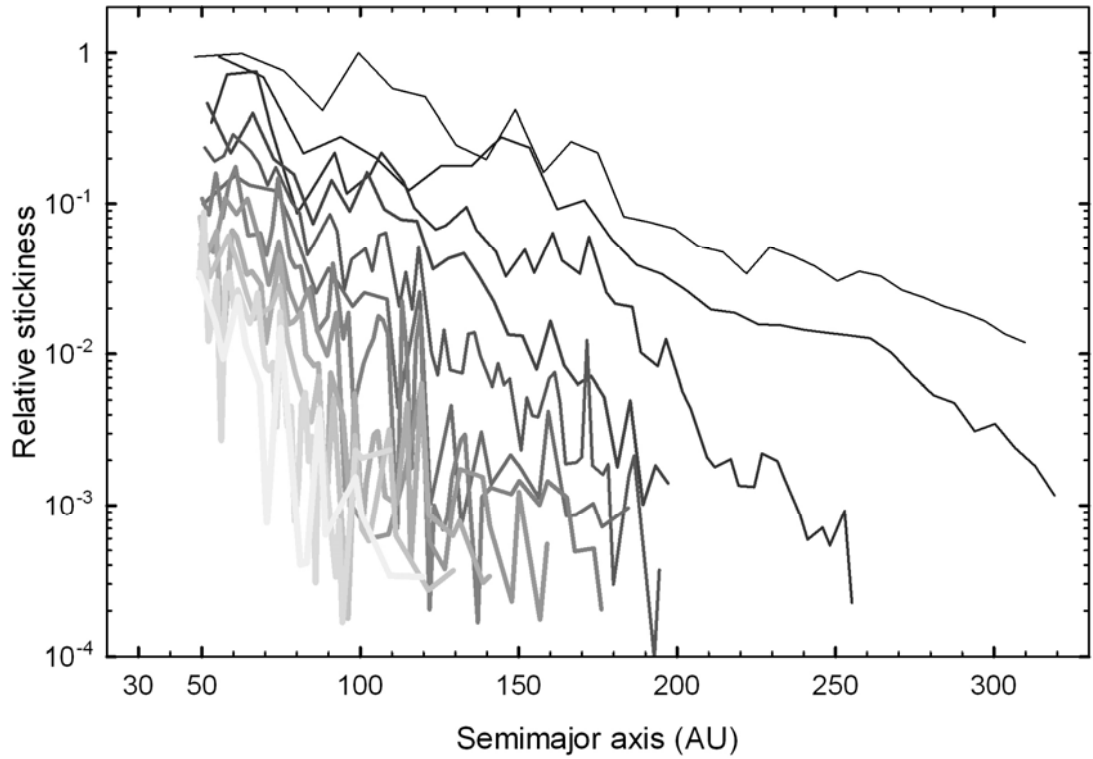

**Figure 5 – Relative resonance stickiness of resonances..., Lykawka, P. S.**

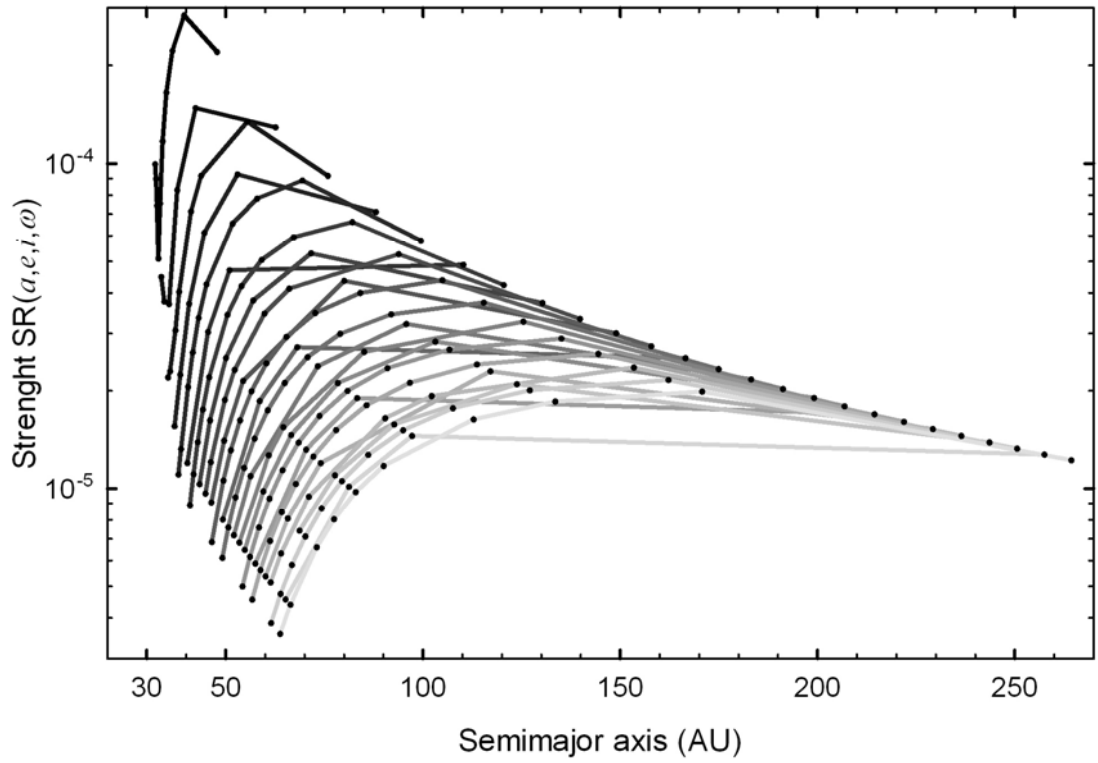

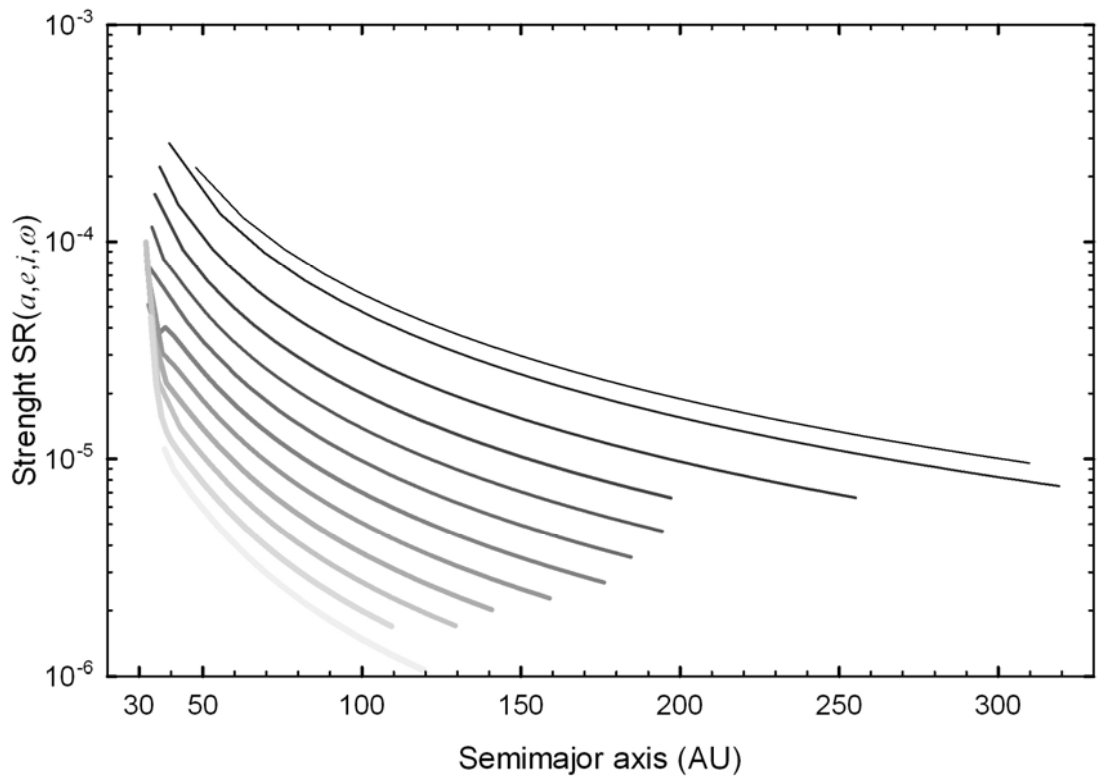

**Figure 6 – Strength of resonances in the scattered disk..., Lykawka, P. S.**

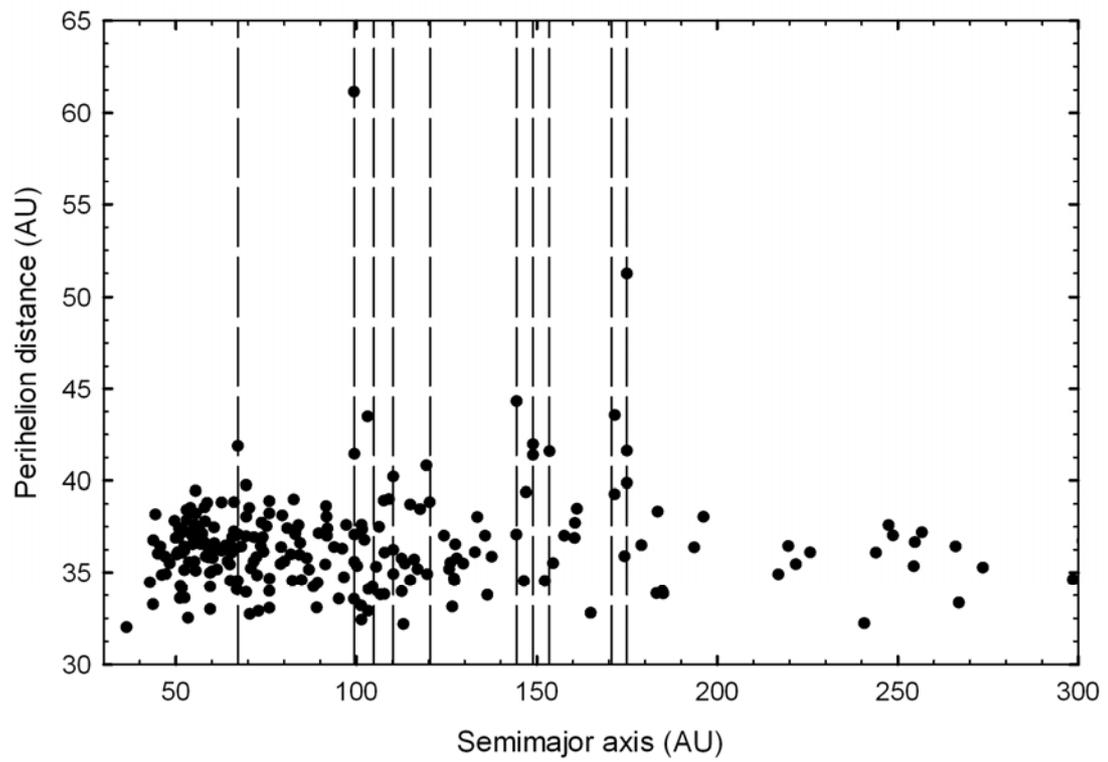

**Figure 7 – Orbital distribution of 255 particles after 4Gyr..., Lykawka, P. S.**